\documentstyle[twoside,fleqn,espcrc2]{article}
\input{psfig}
\newcommand{\pom}{I\!\! P}

\newcommand{\AmS}{{\protect\the\textfont2
  A\kern-.1667em\lower.5ex\hbox{M}\kern-.125emS}}

\begin{document}
\title{Diffraction in hadron-hadron interactions}
\author{K. Goulianos\address{The Rockefeller University,\\
1230 York Avenue, New York, NY 10021, USA\\
(dino@physics.rockefeller.edu)}%
\thanks{Presented at 
``Diffraction 2000, Cetraro, Italy, 2-7 September 2000".}}      

\begin{abstract}
Results on soft and hard diffraction in $pp$ and $\bar pp$ 
collisions are reviewed with emphasis on 
factorization and scaling properties
of differential cross sections. While conventional factorization breaks down
at high energies, a scaling behavior emerges, which leads to a universal 
description of diffractive 
processes in terms of a (re)normalized rapidity gap probability distribution.
\vspace*{-3.5in}
\begin{flushright}
\fbox{RU 00/E-18}\\
\vglue 0.3in
\vglue 0.25cm
\today\\
\end{flushright}
\vspace*{2.6in}
\end{abstract}

\maketitle

\section{INTRODUCTION}
\begin{figure}[htp]
\vglue -0.4in
\centerline{\psfig{figure=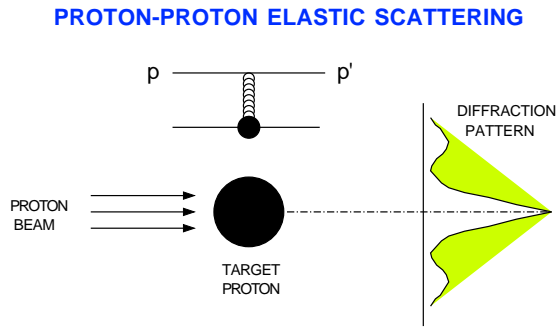,width=3.5in}}
\vglue -3in
\caption{Illustration of diffractive pattern in small angle 
proton-proton scattering}
\label{fig:diff_pattern}
\vglue -0.2in
\end{figure}

The wave nature of particles leads to two classes of diffractive phenomena 
in hadron-hadron collisions: elastic scattering and diffraction 
dissociation. The former, illustrated in Fig.~\ref{fig:diff_pattern}, is 
analogous to the classical diffraction of light. 

\begin{figure}[htp]
\vglue 2.5cm
\centerline{\hspace*{3cm}\psfig{figure=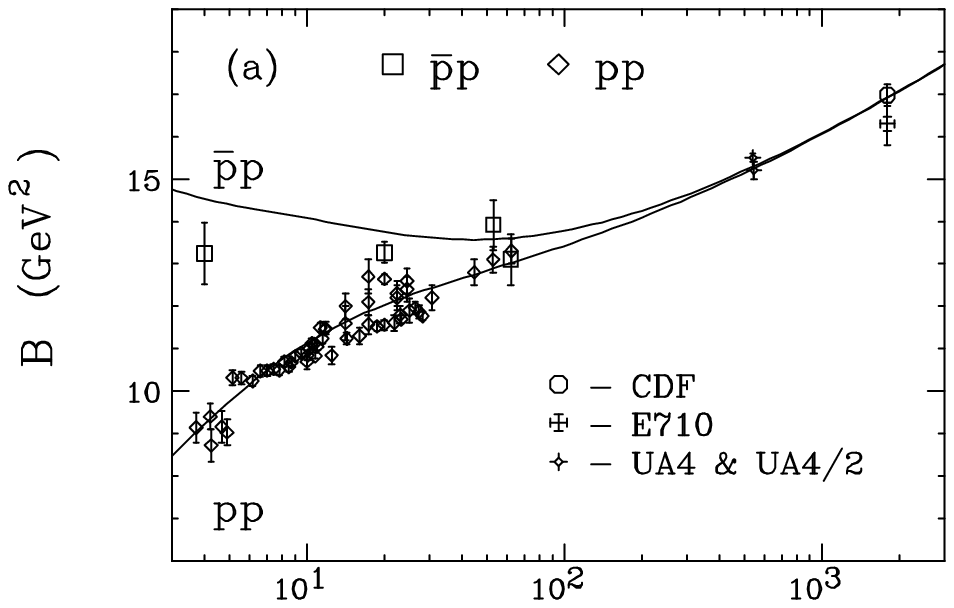,width=6in}}
\vglue -2.3in
\hspace*{1.25in}$\sqrt s$ (GeV)\\
\vglue -1cm
\caption{The slope parameter of $pp$ and $\bar pp$ elastic scattering
in the region of $|t|<0.13$ GeV $^2$~\protect\cite{CMG}.}
\vglue -0.5cm
\label{fig:slope}
\end{figure}

For scattering by a black disc, the differential cross section 
is expected to have the form $d\sigma/dt\sim e^{bt}$, where 
$t=(p'-p)^2\approx -p_T^2$ is the 4-momentum transfer and   
$b$ the {\em slope parameter}. The latter is related to the disc radius,
$R$,  by $b=R^2/4$. Therefore, for a target proton of radius 
$\approx 1/m_{\pi}$,
where $m_{\pi}$ the pion mass, $b$ is expected to have a 
value of  $\frac{1}{4m_{\pi}^2}\approx 13$ GeV$^{-2}$.   
This is indeed approximately what is observed, as shown in Fig.~\ref{fig:slope}.

\begin{figure}[htp]
\vglue -1cm
\centerline{\hspace*{-0.7cm}\psfig{figure=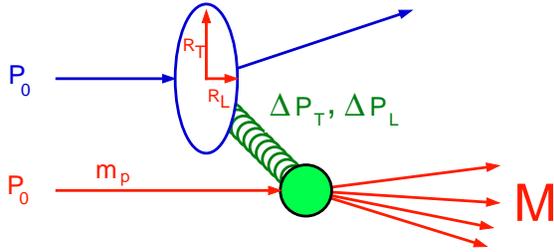,width=4in}}
\vglue -3.9in
\caption{Illustration of diffraction dissociation.}
\vglue -0.2in
\label{fig:sd}
\end{figure}

In contrast to elastic scattering, 
the phenomenon of diffraction dissociation, 
predicted by M.L. Good and W.D. Walker in 1960~\cite{goodwalker},  
has no classical analogue. It can be thought of as the quasi-elastic scattering
between two hadrons, where one of the hadrons is simultaneously excited 
into a higher mass state retaining its quantum numbers.  
This {\em coherent}
excitation, illustrated in Fig.~\ref{fig:sd}, requires not only small 
transverse but also small longitudinal momentum transfer.  
The {\em coherence condition}~\cite{KG} is that the 
longitudinal momentum transfer be smaller than the inverse of the  
longitudinal proton radius, 
$\Delta P_L<\frac{1}{R_L}\approx m_{\pi}\cdot\frac{P_0}{m_p}$.
In terms of the fractional longitudinal momentum loss of the quasi-elastically 
scattered proton, $\xi$, which is related to the diffractive mass $M$
by $\xi\approx M^2/s$, the coherence condition
for diffraction takes the form 
\begin{equation}
\xi\approx \frac{M^2}{s}<\frac{m_{\pi}}{m_{p}}\approx 0.15
\label{eq:coherence}
\end{equation}
The well known large increase of the $d\sigma/d\xi$ distribution in
$pp$ interactions in the region $\xi<0.15$, which approximately 
exhibits the form $\frac{d\sigma}{d\xi}\propto \frac{1}{\xi}$ 
(see Fig.~2 of Ref.~\cite{KG}), 
is testimony to the occurrence of a coherent phenomenon.
 
\begin{figure}[htp]
\vglue -0.4in
\centerline{\psfig{figure=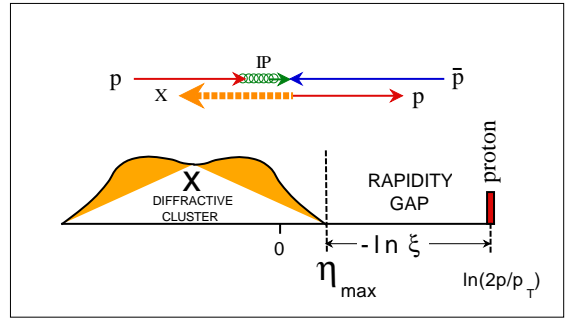,width=4.8in}}
\vspace*{-4.75in}
\caption{Event topology for $p\bar p\rightarrow pX$}
\vglue -0.2in
\label{fig:deltaeta}
\end{figure}

While the wave nature of particles can explain the 
exponential behaviour of the forward (quasi)elastic scattering 
as well as the coherence condition of Eq.~\ref{eq:coherence},
it provides no clue for the 
$1/\xi$ shape of the $d\sigma/d\xi$ distribution. The $\xi$ behaviour 
can be understood in terms of the nature of the exchanged ``particle",
which for diffractive scattering, where no quantum numbers are 
exchanged, must have the quantum numbers of the vacuum. In QCD, this 
``particle", which we will generically refer to here as the 
{\em Pomeron}, is a construct of (anti)quarks and gluons 
in a color singlet state with vacuum quantum numbers. Since such a construct 
does not radiate as it traverses rapidity 
space,~\footnote{We use {\em pseudorapidity} as an approximation to 
{\em rapidity}. Pseudorapidity is defined as $\eta\equiv \ln\frac{2p_L}{p_T}$,
where $p_L$ and $p_T$ are the longitudinal and transverse components of 
the momentum of a 
particle with respect to the beam direction.} a rapidity gap (region of 
rapidity devoid of radiation, i.e. of particles) is associated with the 
exchange of a Pomeron. The width of the rapidity gap, measured from the 
rapidity of the scattered (leading) proton  to that of the emitted 
Pomeron, is given by $\Delta\eta\approx\ln \frac{1}{\xi}$. The event topology
in pseudorapidity space for $p\bar p\rightarrow pX$ is 
shown in Fig.~\ref{fig:deltaeta}.
Since, due to the absence of radiation, 
there is no resistance to the propagation of the Pomeron through 
rapidity space, the cross section should be independent of (or flat in) 
$\Delta\eta$, which through $\Delta\eta=\ln 1/\xi$ leads to 
$d\sigma/d\xi\propto 1/\xi$. 

The Pomeron exchange picture also explains 
why the slope parameter of the $t$ distribution of the leading hadron in 
diffraction dissociation is about one half of that of elastic 
scattering.

\noindent In terms of the form factor of the $\pom pp$ vertex, $F(t)$, the $t$ 
dependence of elastic scattering is expected to be given by 
$F^4(t)\propto e^{b_{el}t}$, while for single diffraction, whose amplitude 
has only one $\pom pp$ vertex, by $F^2(t)\propto e^{b_{sd}t}$, so that  
$b_{sd}=\frac12 b_{el}$.

We have seen that the main features of forward elastic scattering and of
single diffraction dissociation, namely the exponential behaviour 
of the $t$ distributions and the $1/\xi$ dependence, can be understood 
as consequences of coherent scattering resulting from 
the wave nature of particles or, equivalently, from an exchange with 
vacuum quantum numbers. However, there are subtleties in these distributions,
as for example the shrinking of the forward elastic peak with increasing 
c.m.s. energy, whose explanation needs a theoretical framework.
Such a framework has been provided by Regge theory~\cite{Regge}. 
Below, we discuss briefly some Regge theory expectations for 
hadronic diffraction and compare them with experimental results.

\section{THE REGGE APPROACH}
\begin{figure}[htp]
\vglue -0.25in
\centerline{\psfig{figure=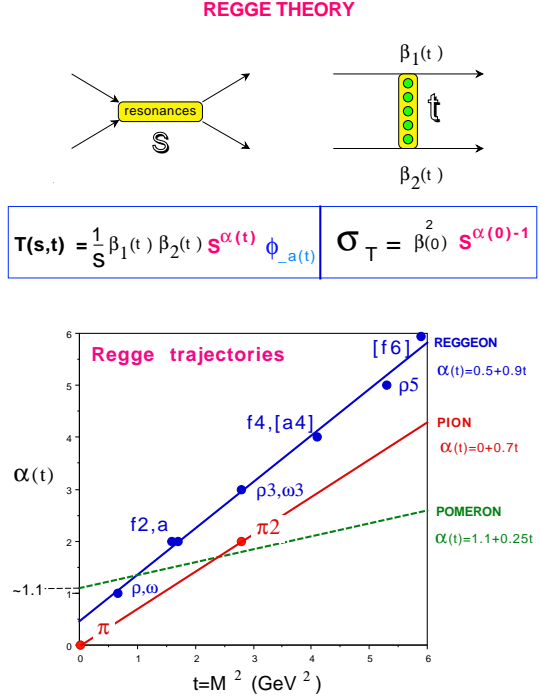,width=3in}}
\vglue -0.3in
\caption{Summary of Regge phenomenology.}
\vglue -0.1in
\label{fig:regge}
\vglue -0.2in
\end{figure}

In the Regge theory approach~\cite{Regge}, 
summarized pictorially in Fig.~\ref{fig:regge},
hadronic interactions are described in terms of $t$-channel exchanges of 
Regge trajectories.  

The three basic Regge trajectories are the Pion, Reggeon and Pomeron, with
intercepts $\alpha(0)$ of approximately 0, 0.5 and 1.1, respectively. 
Because of the $s^{\alpha(t)-1}$ dependence of the amplitude $T(s,t)$ 
in Fig.~\ref{fig:regge}, Pomeron exchange dominates at high  energies.
In fact, the Pomeron trajectory 
with $\alpha(0)\geq 1$ was introduced to account for the 
fact that, at high energies,  hadronic cross sections were found to 
rise with increasing energy, rather than decrease, as would be expected 
from the exchange of the other Regge trajectories.

\begin{figure}[ht]
\vglue -1.7in
\centerline{\hspace*{0.15in}\psfig{figure=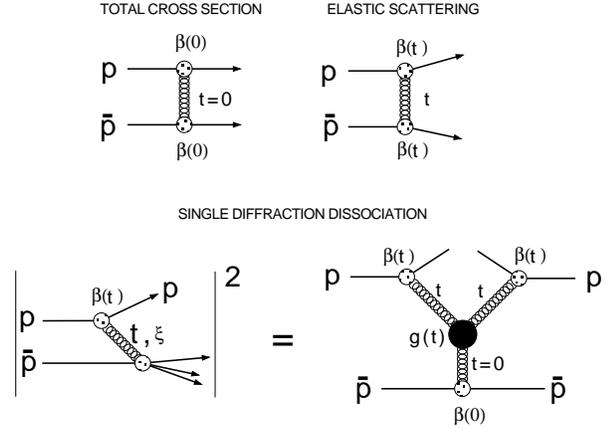,width=4.5in}}
\vglue -2.1in
\caption{Diagrams for total, elastic and single diffraction 
dissociation cross sections.}
\vglue -0.35in
\label{fig:pomeron}
\end{figure}

The Pomeron exchange diagrams for $\bar pp$ interactions
are shown in Fig.~\ref{fig:pomeron}. Through the optical theorem, 
the total cross section is proportional to the $t=0$ elastic scattering 
amplitude. 

The total, elastic and single diffractive cross sections due to 
Pomeron exchange are given by
\begin{equation}
\sigma_T(s)=\beta^2_{\pom pp}(0)
\left(\frac{s}{s_0}\right)^{\alpha_{\pom}(0)-1}
\label{eq:total}
\end{equation}
\begin{equation}
\frac{d\sigma_{el}}{dt}=\frac{\beta^4_{\pom pp}(t)}{16\pi}\;
{\left(\frac{s}{s_0}\right)}^{2[\alpha_{\pom}(t)-1]}
\label{eq:elastic}
\end{equation}
\begin{equation}
\frac{d^2\sigma_{sd}}{d\xi dt}=
\underbrace{\frac{{\beta_{\pom pp}^2(t)}}{16\pi}\;\xi^{1-2\alpha_{\pom}(t)}}_
{f_{\pom/p}(\xi,t)}
\left[\beta_{\pom pp}(0)\,g(t)
\;\left(\frac{s'}{s_0}\right)^{\alpha_{\pom}(0)-1}\right]
\label{eq:diffractive}
\end{equation}

\noindent where $\alpha_{\pom}(t)=
\alpha_{\pom}(0)+\alpha' t=(1+\epsilon)+\alpha' t$
is the Pomeron
trajectory, $\beta_{\pom pp}(t)$ the coupling of the Pomeron to the proton,
$g(t)$ the $\pom\pom\pom$  coupling, $s'=M^{2}$ the $\pom-p$ center of
mass energy squared, $\xi = 1-x_{F}=s'/s=M^2/s$ the fraction of
the momentum of the proton carried by the Pomeron, and $s_0$ an energy
scale parameter traditionally set to the hadron mass scale of 1~GeV$^2$.

Regge theory has been 
shown to provide a good description of experimental 
data in the Fermilab fixed target and ISR energy range 
($\sqrt s<60$ GeV)~\cite{KG}. 
However, as the energy increases, 
the Regge approach becomes infested with unitarity problems, which are 
particularly severe in the case of diffraction dissociation, as discussed 
in the next section.

\begin{figure*}[htp]
\vglue 0.85in
\hbox{{\hspace*{-0.2in}{\psfig{figure=
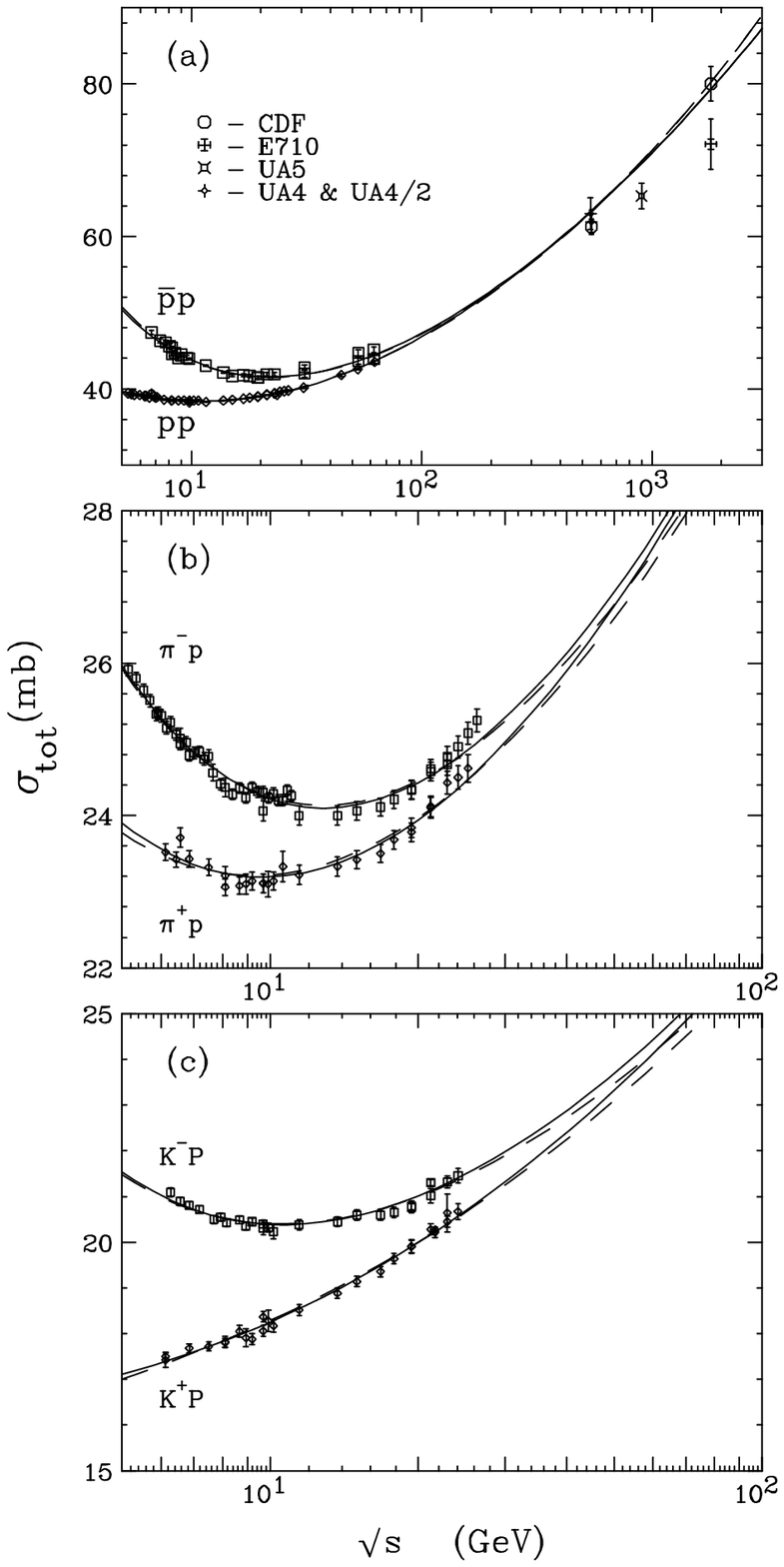,height=3.2in,width=4.8in}}}
{\hspace{-2.4in*}
{\psfig{figure=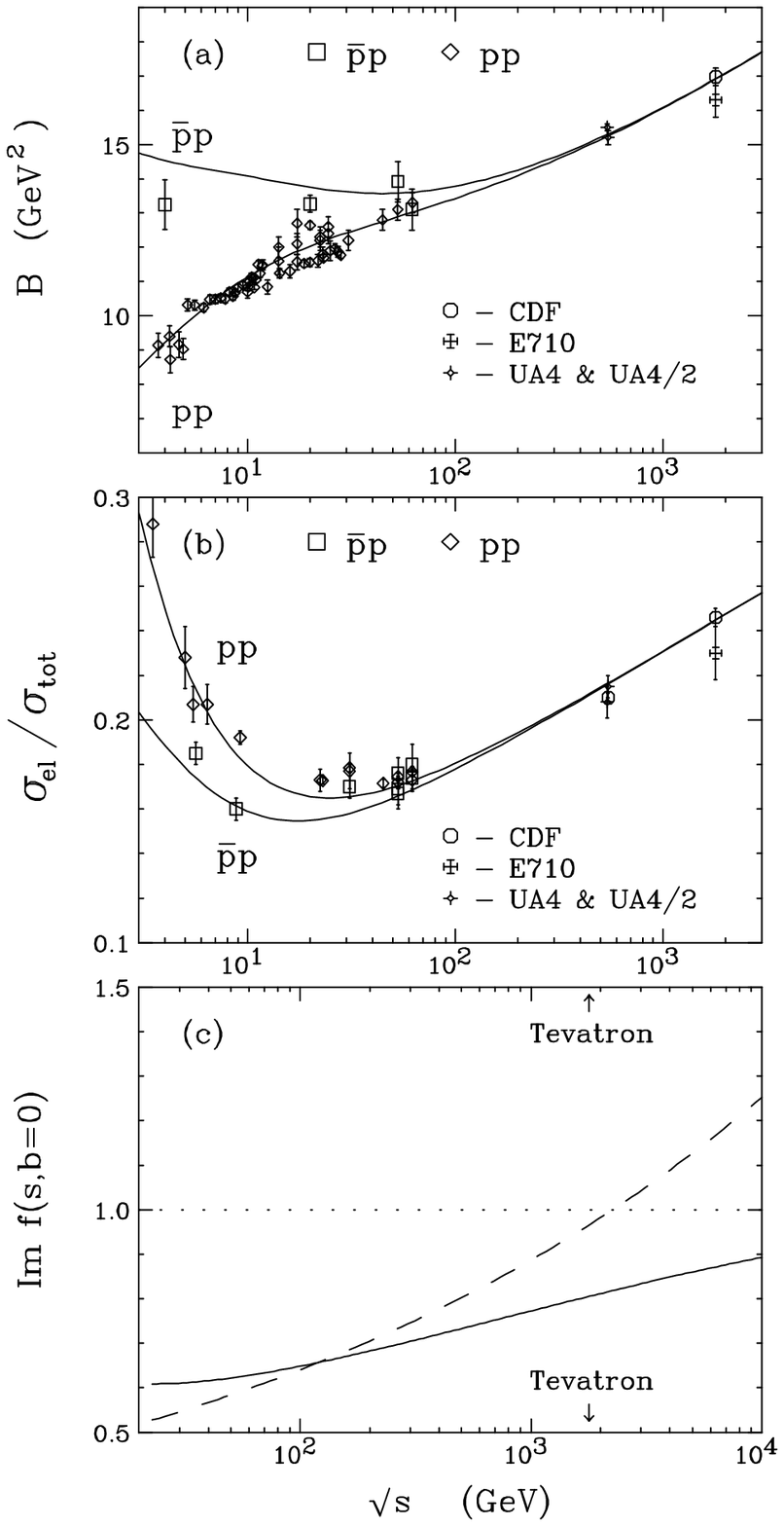,height=3.28in,width=4.92in}}}}
\vglue 1.6in
\caption{({\em left}): $p^\pm p$, $\pi^\pm p$ and $K^\pm p$ 
total cross sections; ({\em right}): 
(a) slope parameter, (b) ratio of elastic to total cross section, and (c)
imaginary part of elastic scattering amplitude in impact parameter space as 
a function of c.m.s. energy for $\bar pp/pp$ interactions.  
The dashed lines are Born level Regge fits,
and the solid lines are fits after eikonilization of the elastic scattering 
amplitude (from Ref.~\protect\cite{CMG}).}
\label{fig:cmg}
\end{figure*}
\section{UNITARITY}
Regge theory with a Pomeron trajectory $\alpha(0)>1$ is plagued 
by unitarity problems as $s\rightarrow \infty$, namely:

(i) The power law $s$-dependence of the total cross section violates the 
Froissart bound~\cite{Froissart}:  
\begin{equation}
\sigma_T\propto s^{\epsilon}>\ln^2s\;{\rm (Froissart\;bound)}
\end{equation}

(ii) The elastic to total cross section ratio increases with $s$ and violates 
the Pumplin bound ($\sigma_{el}<\frac12\sigma_T$):
\begin{equation}
\frac{d\sigma_{el}}{dt}\propto s^{2\epsilon}e^{bt}\;\;\;
(b=b_0+2\alpha'\ln s)
\end{equation}
\begin{equation}
\sigma_{el}\propto s^{2\epsilon}/\ln s\;\;\Rightarrow\;\;
\frac{\sigma_{el}}{\sigma_T}\propto\frac{s^\epsilon}{\ln s}
\end{equation}

(iii) The imaginary part of the forward scattering 
amplitude at zero impact parameter, $Im\,f(s,b=0)$, exceeds unity.

(iv) The ratio of single diffractive to total cross sections
increases with $s$:
\begin{equation}
\frac{d\sigma_{sd}}{d\xi}\propto \frac{1}{\xi^{1+2\epsilon}}\cdot 
(\xi s)^\epsilon
\;\;\Rightarrow\;\;\frac{d\sigma_{sd}}{dM^2}\propto\frac{s^{2\epsilon}}
{\left(M^2\right)^{1+\epsilon}}
\label{eq:scaling}
\end{equation}
\begin{equation}
\frac{\sigma_{sd}}{\sigma_T}\propto s^\epsilon
\end{equation}

In 1992, it was shown~\cite{DL} 
that a good Regge type fit 
to $p^\pm p$, $\pi^\pm p$ and $K^\pm p$ cross sections,
including $\bar pp$ cross sections at $S\bar ppS$ collider energies,  
could be obtained 
using two trajectories, a Pomeron and an {\em effective} Reggeon:
\begin{equation}
\sigma_T^{hp}=Xs^{0.08}+Ys^{0.45}
\label{eq:DL}
\end{equation}
Successful Regge type fits to 
single diffractive differential cross sections 
at Fermilab fixed target and ISR energies had already been obtained 
in 1983~\cite{KG} using a Pomeron trajectory with $\alpha(0)=1$
and an effective Pion trajectory with $\alpha(0)=0$:
\begin{equation}
\frac{d^2\sigma_{sd}}{d\xi\,dt}=\frac{A}{\xi}\,e^{bt}
+B\, \xi\, e^{b't}
\label{eq:pompi}
\end{equation}
It therefore appeared that the unitarity problems inherent in Regge theory 
with Pomeron intercept $\alpha(0)\geq 1$ were not manifest at ``present"
energies. However, the situation changed in 1994 with the CDF measurements of 
elastic, single diffractive and total $\bar pp$ cross sections at 
$\sqrt s=540$ and 1800 GeV~\cite{CDF_edt}. Although good Regge fits 
could still be obtained for elastic and total cross sections, 
two prominent unitarity problems emerged, one in elastic scattering 
and the other in diffraction dissociation. 

In elastic scattering, the Regge prediction for 
the amplitude of the forward elastic scattering in impact parameter space
rises with $\sqrt s$ and exceeds unity at about $\sqrt s=2$ TeV,
violating the unitarity condition $Im\,f(s,b=0)\leq 1$.
This problem can be brought under control 
by eikonalizing the elastic scattering 
amplitude to account for rescattering. Fig.~\ref{fig:cmg} shows Born level 
(dashed) and eikonalized (solid) Regge fits to data~\cite{CMG}.
Both types of fits describe the data well, but as seen in Fig.~\ref{fig:cmg}c 
the extrapolation of the Born level prediction of $Im\,f(s,b=0)$ to energies 
beyond 2 TeV violates unitarity.

\begin{figure}[t]
\vglue -0.75in
\centerline{\psfig{figure=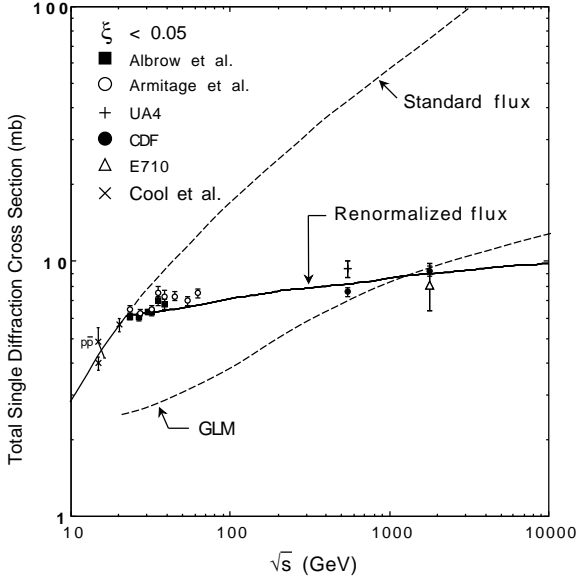,width=3.7in}}
\vglue -1.35in
\caption{The total single diffraction cross section for
$p(\bar p)+p\rightarrow p(\bar p)+X$ versus $\protect\sqrt s$ compared with the
predictions of the renormalized pomeron flux model of
Goulianos~\protect\cite{R}
(solid line)
and of the model of Gotsman, Levin and Maor~\protect\cite{GLM}
(dashed line, labeled GLM); the latter, which includes ``screening
corrections",
is normalized to the
average value of the two CDF measurements at
$\protect\sqrt s=546$ and 1800 GeV.}
\label{fig:GLM}
\vglue -0.25in
\end{figure}

Unitarity was also found to be violated by Regge fits to the $pp/\bar pp$ 
single diffractive cross sections. In their 1994 paper, 
the CDF Collaboration had already 
reported~\cite{CDF_edt} that the $s$ dependence of  
$d\sigma_{sd}/dM^2$ is approximately 
flat between $\sqrt s=20$ and 540 GeV, in contrast 
to the Regge expectation of 
$s^{2\epsilon}$ behaviour (see Eq.~\ref{eq:scaling}).
Eikonalization attempts~\cite{GLM} failed to provide a successful fit to the 
observed $s$-dependence (see dashed line in Fig.~\ref{fig:GLM}).
In 1995, it was proposed~\cite{R} that the ``Pomeron flux", $f_{\pom/p}(\xi,t)$,
represented by the first term in Eq.~\ref{eq:diffractive}, be (re)normalized 
to unity when its integral over all $\xi-t$ space exceeds unity. 
The effect of renormalization is to practically 
cancel out the $s^{2\epsilon}$ dependence in $d\sigma_{sd}/dM^2$,
leading to good agreement with the experimental data (see Fig.~\ref{fig:GLM}). 
The deeper physics meaning of this seemingly ad hoc renormalization proposal 
is discussed in the next section.
\section{A SCALING LAW IN DIFFRACTION}
\begin{figure}[t]
\vglue -0.05in
\centerline{\psfig{figure=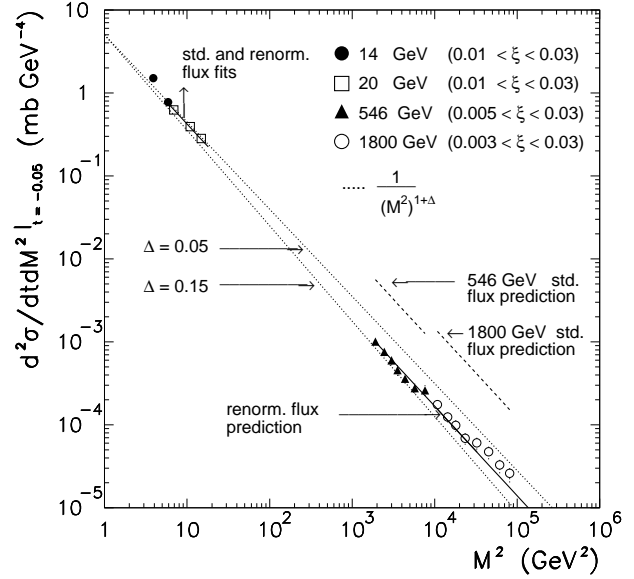,width=3.2in}}
\vglue -0.4in
\caption{Cross sections \protect$d^2\sigma_{sd}/dM^2 dt$
for $p+p(\bar p) \rightarrow p(\bar p)+X$ at
$t=-0.05$ GeV$^2$ and $\protect\sqrt s=14$, 20, 546 and 1800 GeV.
At $\protect\sqrt s$=14 and 20 GeV,
the fits using the standard and renormalized fluxes coincide;
standard (renormalized) flux predictions
are shown as dashed (solid) lines.}
\label{fig:m2}
\vglue -0.25in
\end{figure}

The renormalization of the pomeron flux leads to a scaling behaviour in 
single diffraction, namely the $s$-independence of the $t=0$ differential 
cross section. Figure~\ref{fig:m2} shows $pp$ and $\bar pp$ 
single diffractive cross sections 
at $t=-0.05$ GeV$^2$ as a function of $M^2$ for different $s$ values~\cite{GM}. 
The data have been restricted to $\xi$ regions within which the Pomeron 
contribution dominates and there are no significant distortions from 
$\xi$-resolution effects. The $M^2$ distribution exhibits a 
$1/(M^2)^{1+\epsilon}$ behaviour over the entire $M^2$ region, 
which spans five orders of magnitude. The dotted lines enveloping the 
data represent the 
predictions of the renormalized Pomeron flux model using $\epsilon=0.05$ 
or $\epsilon=0.15$. The data are consistent 
with the same value of $\epsilon$ 
as that extracted from the fit of Ref.~\cite{CMG}
to total and elastic cross sections data, 
namely $\epsilon=0.104$.
The Regge theory predictions for $\sqrt s=540$ and 1800 GeV (dashed lines)
based on extrapolation from $\sqrt s=20$ GeV are significantly 
higher than the data.

The observed scaling behaviour acquires a physical meaning when the 
single diffractive cross section is written in terms of the rapidity gap, 
$\Delta\eta$, rather than the variable $\xi$, using
$\Delta\eta=\ln \frac{1}{\xi}$:
\begin{equation}
\frac{d^2\sigma_{sd}}{dtd\Delta\eta}=
\underbrace{\frac{\beta^2(t)}{16\pi}e^{2(\epsilon+\alpha't)\Delta\eta}}_
{P(\Delta\eta,t)}\;\cdot\;
\underbrace{\kappa\beta^2(0)\left(\frac{s'}{s_0}\right)^
\epsilon}_{\kappa\sigma_T(s')}
\label{eq:sd_prob}
\end{equation}

In the naive parton model, in which the $s^\epsilon$ dependence of the 
total cross section can be understood in terms of 
the total number of the wee partons
~\cite{Levin}, the second term 
in the above equation represents the reduced energy 
($\sqrt s'=M$) cross section, multiplied by a factor 
$\kappa\equiv g(t)/\beta(0)$; the first term may then be interpreted 
as a rapidity gap probability~\cite{R_gap}. Pomeron flux renormalization is 
equivalent to demanding that the integrated gap probability not be 
allowed to exceed unity. In this model, the factor $\kappa$ 
is a color factor introduced to account for the fact that gap formation 
restricts the type of exchanges that lead to the total cross section to 
those of zero net color.

\section{DOUBLE DIFFRACTION}
\begin{figure}[htb]
\centerline{\psfig{figure=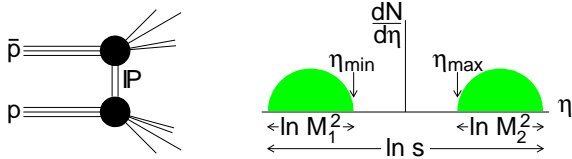,width=3.2in}}
\vglue -2.4in
\caption{Schematic diagram and event topology of a
double diffractive interaction, in which a Pomeron ($I\!\!P$) is
exchanged in a $\bar pp$ collision at center-of-mass energy
$\protect\sqrt s$
producing diffractive masses $M_1$ and $M_2$ separated by a rapidity gap
of width $\Delta\eta=\eta_{max}-\eta_{min}$. The shaded areas represent
regions of particle production.
}
\label{fig:dd} 
\vglue -0.05in
\end{figure}

A stringent test of the normalized gap probability model \cite{R_gap} is
provided by soft double diffraction dissociation, which is illustrated 
in Fig.~\ref{fig:dd}.
 
From Regge theory and factorization, the cross section for double 
diffraction dissociation due to Pomeron exchange has the form
\begin{equation}
\frac{d^3\sigma_{dd}}{dtd\Delta\eta d\eta_0}=
\underbrace{\frac{\kappa\beta^2(0)}{16\pi}e^{2(\epsilon+\alpha't)\Delta\eta}}_
{P(\Delta\eta,\eta_0,t)}\,\cdot\,
\underbrace{\kappa\beta^2(0)\left(\frac{s'}{s_0}\right)^
\epsilon}_{\kappa\sigma_T(s')}
\label{eq:dd_prob}
\end{equation}
\noindent where $\eta_0$ is the center of the rapidity gap, which 
is ``floating" between the two dissociated hadrons, and $\sqrt{s'}$ is
the reduced energy given by 
$\sqrt{s'}=M_1M_2/\sqrt s_0$.

\begin{figure}[htb]
\vglue -2in
\flushright{\Large\bf CDF Preliminary}
\vglue -0.1in
\centerline{\psfig{figure=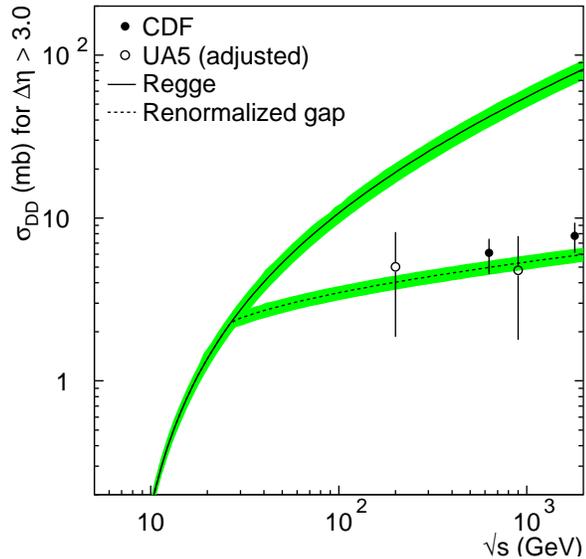,width=3.2in}}
\vglue -0.3in
\caption{The total double diffractive cross section for
$p(\bar p)+p\rightarrow X_1+X_2$ versus $\protect\sqrt s$ compared
with predictions
from Regge theory based on the triple-Pomeron amplitude and
factorization (solid curve) and from the renormalized gap probability model
(dashed curve).}
\vglue -0.2in
\label{fig:ddresult}
\end{figure}

The only $t$-dependence in Eq.~\ref{eq:dd_prob} is due to the term
$e^{2\alpha'\Delta\eta \,t}$, where the rapidity gap is given by 
\begin{equation}
\Delta\eta=\ln\frac{ss_0}{M_1^2M_2^2}
\end{equation}
\noindent Since both nucleons dissociate, there is no 
contribution to the $t$-dependence 
from the nucleon factor factor, as is the case for single 
diffraction. Apart from this difference,
Eqs.~\ref{eq:dd_prob} and ~\ref{eq:sd_prob} for double and single 
diffraction, respectively, are strikingly similar.

The concept of Pomeron flux has no meaning in double diffraction. 
However, in the naive parton model view of diffraction,
the first term in Eq.~\ref{eq:dd_prob} can still be interpreted as a 
rapidity gap probability, while the second term as the reduced energy 
cross section multiplied by {\em the same} color factor $\kappa$ as that 
measured in single diffraction. Thus, 
a comparison of measured double diffractive cross sections to predictions
from Eq.~\ref{eq:dd_prob} with $P(\Delta\eta,\eta_0,t)$ normalized to unity
can provide an unambiguous test of the normalized gap probability model.

In Fig.~\ref{fig:ddresult}, UA5 and preliminary CDF results~\cite{Mary}
on double diffractive cross sections integrated over $t$ and over all 
gaps for $\Delta\eta>3$ are compared with predictions from Eq.~\ref{eq:dd_prob}
without (solid line labeled ``Regge") and with (dashed line labeled 
``renormalized gap") gap renormalization. The data clearly favor the 
renormalized gap model.

\section{THE POMERON}
The introduction of the Pomeron trajectory enabled Regge theory to 
describe the rising total cross sections and the shrinking 
of the forward elastic scattering peak with increasing c.m.s. energy, 
as well as the shape of the
single and double diffraction differential cross sections. However,
a Pomeron with $\alpha(0)\geq 1$ leads to unitarity violations, 
which in single diffraction dissociation are already prominent at 
present accelerator energies. Eikonalization, which brings under 
control the unitarity problem associated with the high energy behaviour of
elastic and total cross sections (see Fig.~\ref{fig:cmg}), 
has not been successful 
in dealing with single diffraction (see Fig.~\ref{fig:GLM}).
Better results have been obtained with Gribov's 
Reggeon calculus approach~\cite{Kaidalov}, 
which involves multi-Pomeron exchange
diagrams, but the associated calculations are cumbersome and difficult 
to implement in hard diffraction processes (discussed in the 
next section). The simplicity of Regge 
theory, which is its strength, is lost in the complexity of the 
remedies proposed to address the unitarity problem. 

In contrast to the difficulties of Regge theory 
associated with unitarity at high energies, 
the data show an amazingly simple and universal $s$-dependence 
in the following two areas:
\begin{itemize}
\item Universality of rising cross sections:
$$\sigma_T\propto s^\epsilon$$
\item Scaling behaviour in diffraction:
$$\frac{d\sigma_{sd}}{dM^2}\propto \frac{1}{\left(M^2\right)^{1+\epsilon}}$$
\end{itemize}
These two scaling laws are the key ingredients used in a new approach 
to diffraction~\cite{parton}, in which the diffractive cross section 
is seen as the reduced energy parton model total cross section multiplied 
by a color factor and a normalized rapidity gap probability.

\section{HARD DIFFRACTION}

Hard diffraction processes are hadronic interactions
incorporating a high transverse momentum partonic scattering
while carrying the characteristic signature of diffraction, namely
leading beam particles and/or large rapidity gaps. As in soft diffraction, 
such processes are believed to be mediated by Pomeron exchange. 
The generic QCD view of the Pomeron is a gluon/quark color-singlet state
with vacuum quantum numbers. A question of great theoretical
interest is whether the Pomeron has a unique particle-like partonic structure.
This question can be addressed experimentally by studies of structure functions 
in events with a diffractive signature~\cite{IS}.

\begin{figure}[h,t]
\vglue -0.5in
{\hspace*{-0.5in}\psfig{figure=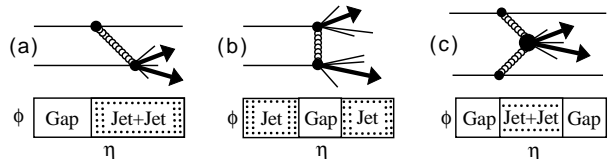,width=4.2in}}
\vglue -5in
\caption{Dijet production diagrams and event topologies for
(a) single diffraction (b) double diffraction
and (c) double Pomeron exchange.}
\label{topology}
\end{figure}

In hadron-hadron interactions, 
there are three types of hard diffraction processes
accessible to experimentation with present day accelerators:
single diffraction (SD), double diffraction (DD) and double pomeron
exchange (DPE). The event topology of dijet events produced in
these processes is shown, respectively,
in Figs.~\ref{topology}(a), (b) and (c). All three processes can be tagged by
the rapidity gap signature. Single diffraction and DPE can also be tagged
by detecting the leading particle(s) on the gap side.
\vglue -0.3in

The first observation of a hard diffractive process was made by the 
UA8 Collaboration at the $S\bar ppS$ collider 
in a study of dijet events 
produced  in association with a leading proton~\cite{UA8}.
Using rapidity gap tagging, the CDF and D\O\, Collaborations have subsequently
studied dijet production in all three processes shown in Fig.~\ref{topology}. 
In addition, CDF has studied diffractive $W$,
$b$-quark and $J/\psi$ production, as well as dijet production
in SD and DPE using a ``roman pot" magnetic spectrometer to detect
leading antiprotons (in the DPE study the events were tagged by
a leading antiptoton and a rapidity gap on the proton side).

\begin{table*}[t]
\caption{Diffractive to total 
production ratios at the Tevatron.}
{\begin{tabular}{|l|c|c|c|c|}
\hline
&&&&\\
Hard process&$\sqrt{s}$ (GeV)&$R=\frac{\rm DIFF}{\rm TOTAL}\,(\%)$&
Comments&Exp't\\
&&&&\\
\hline
{\fbox{SD}}&&&&\\
$W(\rightarrow e\nu)$+G&1800&$1.15\pm 0.55$&$E_T^e,\;/\!\!\!\!E_T>20$ GeV
&CDF~\cite{CDF_W}\\
Jet+Jet+G&1800&$0.75\pm 0.1$&$E_T^{jet}>20$ GeV, $\eta^{jet}>1.8$
&CDF~\cite{CDF_JJG}\\
$b(\rightarrow e+X)$+G&1800&$0.62\pm 0.25$&$|\eta^e|<1.1$, $p_T^e>9.5$ GeV
&CDF~\cite{CDF-b}\\
\hline
\fbox{DD}&&&&\\
Jet-G-Jet&1800&$1.13\pm 0.16$&$E_T^{jet}>20$ GeV, $\eta^{jet}>1.8$
&CDF~\cite{CDF_JGJ1800}\\
Jet-G-Jet&1800&$0.54\pm 0.17$&$E_T^{jet}>12$ GeV, $\eta^{jet}>1.6$
&D\O\,~\cite{D0_JGJ}\\
Jet-G-Jet&630&$1.85\pm 0.37$&$E_T^{jet}>12$ GeV, $\eta^{jet}>1.6$
&D\O\,~\cite{D0_JGJ}\\
Jet-G-Jet&630&$2.7\pm 0.9$&$E_T^{jet}>8$ GeV, $\eta^{jet}>1.8$
&CDF~\cite{CDF_JGJ630}\\
\hline
\end{tabular}

\label{table:fractions}
}
\end{table*}

The published diffractive to non-diffractive 
ratios~\cite{CDF_W}-\cite{CDF_JGJ630}
obtained in the studies 
using rapidity gap tagging are presented in Table~\ref{table:fractions}.
Both the SD and DD fractions are $\approx 1\%$ at $\sqrt s=1800$ GeV 
and $2-3\%$ at $\sqrt s=630$ GeV. 
\begin{itemize}
\item The process independence of the diffractive fractions at a given 
energy shows that the partonic structure of the Pomeron, and in particular the 
gluon to quark content,  
is not very different from that of the proton. 
\item The increase of the DD fraction with decreasing energy 
follows the $s^{-2\epsilon}$ dependence expected from the rapidity gap 
(re)normalization factor. With $\epsilon\approx 0.2$, which is the value 
measured in diffractive DIS at HERA, the 630 to 1800 GeV ratio is
predicted to be $(630/1800)^{-4\epsilon}=2.3$, in agreement with the 
CDF and D\O\, results.
\end{itemize}

\begin{figure}[hp]
\vglue -1cm
\centerline{\psfig{figure=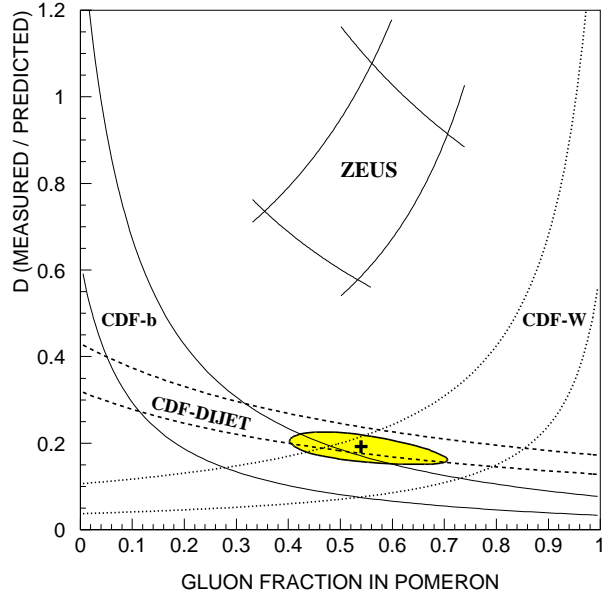,width=3.5in}}
\vglue -1cm
\caption{The ratio, $D$, of measured to predicted diffractive
rates as a function of the gluon content of the Pomeron.
The predictions are from POMPYT using the standard Pomeron flux
and a hard Pomeron structure.
The CDF-$W$ curves were calculated
assuming a three-flavor quark structure for the Pomeron.
The black cross and shaded ellipse
are the best fit and $1\sigma$ contour of a least square two-parameter fit
to the three CDF results.}
\vglue -1cm
\label{fig:f_g}
\end{figure}

The gluon fraction of the Pomeron was measured by CDF by combining the 
diffractive dijet, $W$, and $b$-quark measurements. 
Assuming the standard Pomeron flux in the POMPYT Monte Carlo 
program~\cite{POMPYT},
the ratios $D$ of measured to POMPYT-predicted
SD to ND fractions of $W$, dijet, and $b$-quark production rates
trace  different curves in the plane of $D$ versus $f_g$.
Figure~\ref{fig:f_g}
shows the $\pm1\sigma$ curves corresponding to the results.
From the oval-shaped overlap of the $W$, dijet and $b$-quark
curves (shaded area),  CDF obtained $f_g=0.54^{+0.16}_{-0.14}$ and 
$D=0.19\pm 0.04$.
The decrease of the value of 
$D$ from HERA to the Tevatron
represents a breakdown of factorization.

\section{THE DIFFRACTIVE STRUCTURE FUNCTION OF THE NUCLEON}
In $\bar pp$ collisions, the diffractive structure function (DSF)
of the (anti)proton is defined in the same manner as the non-diffractive SF,
except that in addition to being a function of $x$ and $Q^2$ it is also 
a function of $\xi$. The DSF was measured by CDF using a sample of 
diffractive dijet events tagged by a leading antiproton~\cite{CDF_JJ}. 
Another event sample consisting of dijet events collected with a minimum
bias trigger was used for monitoring. 
The procedure followed is based on measuring the
ratio $R(x)$ of SD to ND cross sections as a function of the Bjorken-$x$ of the
parton in the $\bar p$ participating in the hard scattering. In LO QCD,
this ratio is proportional to the corresponding structure functions.
The DSF is obtained by multiplying the measured $R(x)$ by the known
ND structure function.

\begin{figure}[ht]
\vglue -1.7cm
\centerline{\psfig{figure=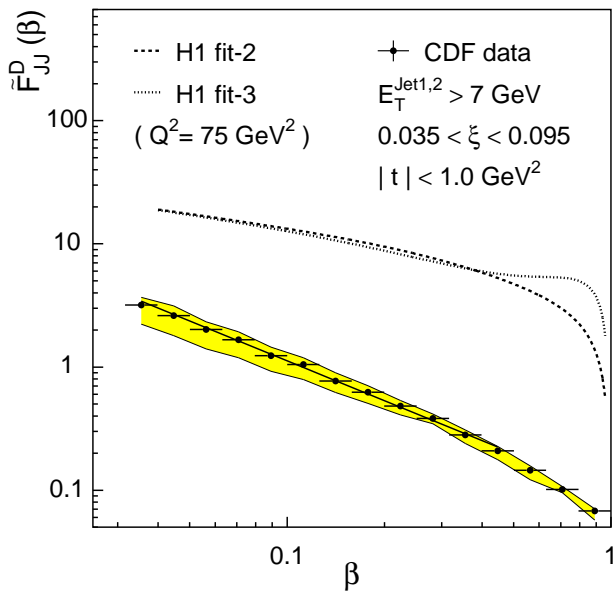,width=3.5in}}
\vglue -1cm
\caption{Data $\beta$ distribution (points) compared with expectations from
the parton densities of the proton extracted from diffractive deep
inelastic scattering by the H1 Collaboration at HERA.
}
\label{fig:sd4}
\vglue -1cm
\end{figure}

The value of $x$ of the parton in the $\bar p$ was evaluated from the jet 
$E_T$ and $\eta$ values, as follows:
$$x=\frac{1}{\sqrt{s}}\sum_{i=1}^nE_T^ie^{-\eta^i}$$
The sum was carried out
over the two leading jets plus the next highest
$E_T$ jet, if there was one with $E_T>5$ GeV.
The structure function relevant to dijet production is a color-weighted
combination of quark and gluon components:
$$F_{jj}(x)=x\left\{g(x)+\frac49\sum_i [(q_i(x)+{\bar q}_i(x)]\right\}$$
\noindent where $g(x)$ and $q(x)$ are gluon and quark parton densities,
respectively. For comparisons with predictions based on HERA results,
in which the DSF is usually presented in terms of the variable $\beta$
instead of $x$ ($\beta\equiv x/\xi$ may be interpreted as the 
momentum fraction of the parton in the Pomeron), 
the DSF obtained from the equation
$F^D_{jj}(x,\xi)=R(x,\xi)\times F^{ND}_{jj}(x)$ was transformed to
${F}^D_{jj}(\beta,\xi)$ by a change of variables.
The resulting ${F}^D_{jj}(\beta,\xi)$ is presented in Fig.~\ref{fig:sd4}
as a function of $\beta$ along with expectations based on
diffractive parton densities extracted by the H1 Collaboration from
diffractive DIS measurements. The CDF measured 
$\tilde{F}_{jj}(\beta)$ (the tilde denotes integration over the indicated 
$\xi$ and $t$ ranges) differs from the prediction based on HERA data 
both in shape and normalization. The normalization discrepancy 
is of ${\cal{O}}(0.1)$,
confirming the breakdown of factorization observed in the comparison of the
rapidity gap results with expectations from HERA measurements
(see Fig.~\ref{fig:f_g}).

\section{DOUBLE POMERON DIJETS}
\begin{figure}[htp]
\vglue -1cm
\centerline{
\hspace{0.3cm}\psfig{figure=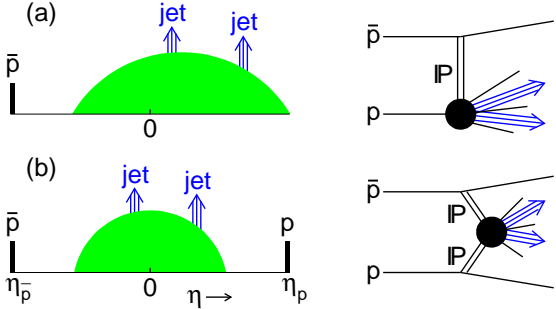,width=3.25in}
}
\vglue -1.8in
\caption{Illustration of event topologies in pseudorapidity, $\eta$,
and associated Pomeron exchange
diagrams for dijet production in (a) single diffraction and (b) double
Pomeron exchange. The shaded areas on the left side represent
particles not associated with the jets (underlying event).}
\label{fig:dpe1}
\vglue -0.25in
\end{figure}

An interesting test of factorization has been performed by CDF 
by comparing the diffractive structure function measured in SD to 
that measured from dijet production in DPE.
The DPE process is illustrated in Fig~\ref{fig:dpe1}b. The DPE signal 
was extracted from the roman pot diffractive dijet event sample 
by requiring a rapidity gap (RG) on the proton side.

\begin{figure}[t]
\vglue -0.7cm
\centerline{\psfig{figure=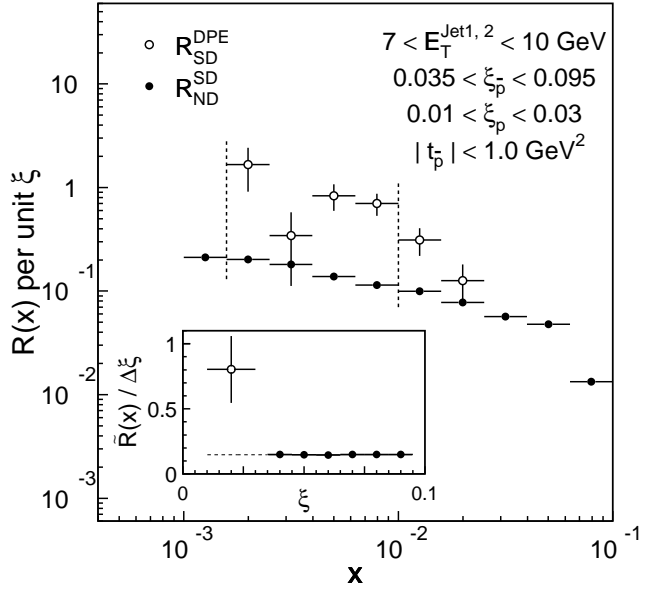,width=3.5in}}
\vglue -1cm
\caption{Ratios of DPE to SD (SD to ND) dijet event rates per
unit $\xi_p$ ($\xi_{\bar{p}}$), shown as open (filled) circles,
as a function $x$-Bjorken of partons
in the $p$ ($\bar{p}$).
The errors are statistical only. The SD/ND ratio has a normalization
systematic uncertainty of $\pm 20\%$.
The insert shows $\tilde{R}(x)$ per unit $\xi$ versus $\xi$,
where the tilde over the $R$ indicates the weighted average of
the $R(x)$ points in the region of $x$ within the vertical dashed lines,
which mark the DPE
kinematic boundary (left) and the value of
$x=\xi_p^{min}$ (right).}
\label{fig:dpe4}
\vglue -1cm
\end{figure}

In events with a leading antiproton (LA), or equivalently with a rapidity gap 
on the $\bar p$ side, 
the ratio of the DPE to SD dijet production cross sections at the same 
$x_p$ for fixed $\xi_p$, $R^{DPE}_{SD}(x_p,\xi_p)$, 
is in LO QCD equal to the ratio of the 
SD to ND structure functions of the proton. 
Therefore, diffractive factorization can be tested by comparing this ratio with 
the SD to ND ratio, $R^{SD}_{ND}(x_p,\xi_p)$, for SD events  
with no rapidity gap on the antiproton side.
Since no such events were available, the comparison was made with 
the measured ratio $R^{SD}_{ND}(x_p,\xi_{\bar p})$.
The result is shown in Fig.~\ref{fig:dpe4}.
The vertical dashed lines mark the DPE
kinematic boundary (left) and the value of
$x=\xi_p^{min}$ (right).
The weighted average of the DPE/SD points in the region
within the vertical dashed lines is $\tilde{R}^{DPE}_{SD}=0.80\pm 0.26$.

Factorization demands that $\tilde{R}^{DPE}_{SD}$ be the same
as $\tilde{R}^{SD}_{ND}$ at fixed $x$ and $\xi$.
Since the $\xi_p$ and $\xi_{\bar p}$ regions, which are respectively relevant
for the DPE/SD and SD/ND ratios, do not overlap,
the $\xi$ dependence of the ratios $\tilde{R}(x)$ (per unit $\xi$),
where the tilde over the $R$  indicates
the weighted average of the points in the region of $x$
within the vertical dashed lines in the main figure, 
was examined and found to be flat in $\xi$ (see inset of Fig.~\ref{fig:dpe4}).
A straight line fit to the
six $\tilde{R}^{SD}_{ND}$ ratios extrapolated to $\xi=0.02$
yields $\tilde{R}^{SD}_{ND}=0.15\pm 0.02$.
The ratio of $\tilde{R}^{SD}_{ND}$ to $\tilde{R}^{DPE}_{SD}$
is $D\equiv \tilde{R}^{SD}_{ND}/\tilde{R}^{DPE}_{SD}=0.19\pm 0.07$.
The deviation of $D$ from unity represents a breakdown of factorization.

In Fig.~\ref{fig:dpe1}, the presence of the rapidity gap on the antiproton side 
reduces the rapidity range over which a gap can be formed on the proton side.
Thus, $D$ decreases
as the $\eta$-range available for the
formation of a rapidity gap increases. 
This behaviour is in accordance with the (re)normalized gap probability 
predictions~\cite{R,R_gap}.

\section{CONCLUSIONS}
The central issue in hadronic diffraction is the question of universality 
of the rapidity gap probability. Another important issue is that of the 
existence of a unique, process independent diffractive structure function.

Soft single diffractive $pp$ and $\bar pp$ data at low $\xi$ and $t$ 
have successfully been described~\cite{GM} by the product of two terms, one 
proportional to the total cross section at the reduced c.m.s. energy,
$\kappa\sigma_T(s')$, and the other representing a normalized rapidity gap 
probability, $P(\Delta\eta,t)=P_0e^{2(\epsilon+\alpha't)\Delta\eta}$.
Recent CDF data on double diffraction dissociation support this 
description~\cite{Mary}.
Comparisons of hard diffraction results with POMPYT Monte Carlo predictions 
also show good agreement with the gap probability (re)normalization hypothesis.
Thus, the data show a remarkable universality in rapidity gap formation
extending across soft and hard processes.

The observed process independence of hard SD to ND 
ratios (see table~\ref{table:fractions})
indicates that the partonic composition of the Pomeron 
is similar to that of the proton. The discrepancy in shape and normalization 
between the measured DSF at the Tevatron and expectations based on 
HERA measurements (Fig.~\ref{fig:sd4}) represents a breakdown of factorization.
A normalization discrepancy has also been found 
between the DSF's measured in SD and DPE at the Tevatron. 
The observed discrepancy is 
foreseen in the RG (re)normalization model~\cite{R,R_gap}.

\end{document}